\documentclass[amscd,amssymb,verbatim]{amsart}
\usepackage{graphics}
\begin{document}
\newcommand{\ssum}[1]{\mbox{\footnotesize$\displaystyle\sum_{{#1}}$}}
\title[Minimax Games and Spin Glasses]%
  {Minimax Games, Spin Glasses and the Polynomial-Time Hierarchy of
  Complexity Classes}
\author{Varga Peter}
\address{Institute of Mathematics and Informatics \\
        Lajos Kossuth University \\
        H-4010 Debrecen, Pf.12. Hungary}
\email{varga@@math.klte.hu}
\date{}

\begin{abstract}
We use the negative replica method, which was originally developed for the
study of overfrustation in disordered system, to investigate the statistical
 behaviour of the cost function of
minimax games. These games are treated as hierarchical statistical
mechanical systems,    in which one of the components is at negative
 temperature.
\end{abstract}

\maketitle

{\em (PACS 61.43Fs, 64.60Cn, 89.90+h)}

\section{Introduction} \label{S:intro}

The theory of spin glasses has found interesting applications in
several branches of science \cite{mpv}. In the theory of
combinatorial optimization it inspired the invention of the so called
simulated annealing heuristic search technique \cite{simann}. With the
help of the replica method, several authors \cite{mp,fuan,mp2} managed to
 obtain analytical insight about  optimal solutions of some
hard optimization problems.  The interest in these studies was driven
by the fact that many of these problems were members of the {\bf NP}
complexity class, which means that to check their solutions requires only
polynomial time, but to find them is presumably much harder. 

{\bf NP} is among the first few members of a hierarchy of 
 complexity classes of increasing difficulty, the polynomial-time
 hierarchy {\bf PH} 
of Meyer and Stockmeyer \cite{book}. Examples of problems from this
hierarchy are adversary games, where the first player tries
to minimize the objective function while the second one tries to
maximize it. In this paper we treat the case in which one of the
players has control over the spins of a spin glass, while the other
controls the external magnetic field, and the
objective function is the energy of the spin configuration.

The standard machinery of statistical mechanics provides information
on the ground state (minimum of energy), as the temperature approaches
zero. To study the maximum, we need to approach zero from 
negative direction. Fortunately, this step can be incorporated into the
replica method by allowing the number of replicas to be
negative. The method of negative replicas was invented by Dotsenko, Franz
 and Mezard  to study partial annealing and overfrustation in disordered
 systems \cite{DFM}. (Some related works are \cite{D,PCS,CPS,FD}.)
 We use this framework for the investigation of minimax games.

In Sec. II we give a short, nontechnical description of the
polynomial-time hierarchy of complexity classes. In Sec. III we apply
this  extension of the replica method for three simple models. Sec. IV 
contains an extension  of the negative replica method  for multi-move games.

\section{The polynomial-time hierarchy} \label{S:pol}

In this section we closely follow the exposition of Stockmeyer
\cite{book}. To formulate a rough definition of the complexity classes
it is easier to use decision problems than optimization ones. We
define the complexity class {\bf P} as those problems which are
solvable by a deterministic (and sequential) computer in time bounded
by some polynomial of the size of the problem. Of course, one should
spell out in a little more detail the kind of computers used
(usually a Turing or Random Access Machine), however, the class {\bf
P} is remarkably stable with respect to changes of the
computational model. 

The definition of the class {\bf NP} is  similar, but in this case  the
use of nondeterministic computers is allowed.  The nondeterministic
model of computation is more powerful than the deterministic one. Let
us take for example the most representative problem of {\bf NP}, the
satisfiability of an arbitrary Boolean expression. If there is an
assignment of truth values to the variables of the expression such that the
 expression evaluates to 'true', then a
nondeterministic computer is able to verify that in polynomial time. In
the first few steps it correctly guesses that assignment, and then by
a deterministic algorithm it verifies that the assignment indeed
satisfies the Boolean expression.  These steps take only polynomial
time. So {\bf NP} contains those problems, whose solutions, if exist,
can be checked in polynomial time. A basic conjecture of computer
science is that the inclusion ${\bf P} \subset {\bf NP}$ is proper,
i.e. there are problems easy to check but hard to solve.

In the case of spin glasses the decision problem is that given a
$J_{ij}$ coupling constants matrix and a number $K$, is there any spin
configuration $s_{i}$ such that $E_{J}(s_{i})= \sum_{i,j} J_{ij}s_{i}s_{j}
\leq K$. (To keep the size of the problem under control, $J_{ij}$
should take only discrete  (maybe $\pm 1$) values).  A closely related
problem class is {\bf co-NP}, the complement of {\bf NP}. Here the
task is to recognize those problems which has  no solution. For
example in the spin glass case one needs to prove that there is no
spin configuration with energy less than a given constant.  It is
unlikely that such proof of polynomial length exists for a random
$J_{ij}$ matrix , so it is believed that {\bf NP} $\neq$ 
{\bf co-NP}.

Optimization problems requires the ability to solve both {\bf NP} and
{\bf co-NP} problems. To prove that $U_{0}$ is the minima of
$E_{J}(s)$, one should find first $s$ such that $U_{0}=E_{J}(s)$, then
solve a {\bf co\mbox{-}NP} problem proving that there is no such $s$ that
$E_{J}(s) <U_{0}$.

Several ways exist to obtain problems harder than {\bf NP}. The most
obvious is to allow more (say exponential) time for the computation. A
more subtle way to increase the power of the computational model is
the use of oracle machines.  They have an additional instruction
'Call-Oracle' . When the machine executes this instruction, it
presents the oracle a problem from the oracle's problem class
for which the  oracle gives returns the solution or gives 'no'
 answer in a single step.
The power of an oracle computer depends on the oracle's problem class
$C$. Since the oracle recognizes non-membership in $C$, too, the
oracles $C$ and $co$-$C$ have the same computational power. In this
manner, {\bf NP($C$)} (resp. {\bf P($C$)}) is defined as the decision
 problem
class, which satisfiability can be decided by a nondeterministic
(resp. deterministic) computer with oracle $C$ in polynomial time.

By denoting {\bf P}=
$\Sigma_{0}^{P}$, the polynomial-time hierarchy is
defined as 
\[ 
\Sigma_{k}^{P}={\bf NP}(\Sigma_{k-1}^{P} ),\quad 
\Delta_{k}^{P}={\bf P}(\Sigma_{k-1}^{P} ),\quad
\Pi_{k}^{P}=co\mbox{-}\Sigma_{k}^{P}.
\]
Members of this hierarchy occurs in problems involving the alternation
of existential and universal quantifiers. The satisfiability of the
Boolean formula $f({\bf x})$ (i.e. $f \in {\bf NP}$) means
$\exists{\bf x}f({\bf x})$, while its non-satisfiability (i.e. $f \in
{\bf co\mbox{-}NP}$) is the same as $\forall{\bf x}\neg f({\bf x})$.  Boolean
formulas $\exists{\bf x}_{1}\forall{\bf x}_{2}...\exists{\bf x}_{2l+1}
f({\bf x}_{1},{\bf x}_{2},...)$ or \\  
$\exists{\bf x}_{1} \forall{\bf x}_{2}...\forall{\bf x}_{2l}
\neg f({\bf x}_{1},{\bf x}_{2},...)$ with $(k-1)$-fold alternation of
existential and universal quantifiers provides natural examples for
problems from $\Sigma_{k}^{P}$.  The determination of the
satisfiability of such formulas can be described as a game between two
adversary players. The first player's objective is to satisfy the
formula, while the second one tries to set the variables ${\bf x}_{2}$,
${\bf x}_{4}$,.., so that the formula is not satisfied.  

An optimization problem from the polynomial-time hierarchy is the
determination of the outcome 
\[
M=\max_{{\bf x}_{1}}\min_{{\bf x}_{2}}...c({\bf x}_{1},{\bf
x}_{2}...)
\]
of a minimax game. For many functions $c({\bf x}_{1},{\bf x}_{2}...)$,
the computation of $M$ is a $\Delta_{k+1}^{P}$ type problem if there
are $k-1$ alternation of the the $min$ and $max$ operators. In the
next section we treat the case where ${\bf x}_{1}$ and ${\bf x}_{2}$
represent sets of discrete spin variables and $c({\bf x}_{1},{\bf
x}_{2})$ is the  energy function of spin configurations.

\section{Spin games} \label{S:game}

In this section we study two-move minimax games. The objective function
is denoted by $H(u,v)$, where $u$ and $v$ are two sets of
variables. The first player (the minimizer) controls the $u$
variables, while the second one (the maximizer) controls the $v$
variables. If both players play optimally. then the outcome of the
game is 
\[  
M=\inf_{u}\left(\sup_{v}H(u,v)\right).
\]
To apply the methods of statistical mechanics, $\inf_{u} h(u)$ (resp.
$\sup_{v} H(u,v)$) is replaced  by the free energy of a system with
Hamiltonian $h$ (resp. $H$) at low positive (resp. negative)
temperature. For that purpose we introduce
\begin{equation}
  \label{E:Mbb}
  \begin{aligned}
    M(\beta_{u},\beta_{v}) &=-\frac{1}{\beta_{u}} \ln 
    \sum_{\{u\}}\exp -\beta_{u}
    \left(\frac{1}{\beta_{v}} \ln \sum_{\{v\}} \exp \beta_{v}
      H(u,v) \right)  \\
    &= -\frac{1}{\beta_{u}} \ln \sum_{\{u\}} \left( \sum_{\{v\}} \exp
      \beta_{v} H(u,v) \right)^{-\beta_{u} / \beta_{v}} \\
    &= -\frac{1}{\beta_{u}} \lim_{n \to 0} \frac{1}{n}
    \left[ \left( \sum_{ \{ u^{a},v^{\alpha} \} }  \exp
        \beta_{v} \sum_{ \{ a,\alpha \} } H(u^{a},v^{\alpha}) 
      \right)-1 \right].
  \end{aligned}
\end{equation}
There are $n$ replicas of $u$ and
$nk=-n\beta_{u}/\beta_{v}$ copies of $v$. If $\beta_{u},\beta_{v} \to
\infty$, then $M(\beta_{u},\beta_{v}) \to M$. (At least if the zero
temperature entropy vanishes, which is true even in the mean field
theory of spin glasses.)

To gain some experience with the method of negative replicas, we
apply it first for the non-random Hamiltonian
\begin{equation}\label{E:j1}
  H(u,v)=\frac{2}{N}
  \left(
    \ssum{i}u_{i}
  \right)
  \left(
    \ssum{i}v_{i}
  \right)
  +g \ssum{i}u_{i} +h \ssum{i}v_{i}, \qquad{} i=1..N.
\end{equation}

In this case the application of the $n \to 0$ limit is not
necessary, so the $u$ spins are not replicated. The partition function
is
\begin{multline}\label{E:j2}
  Z=\sum_{u,v^{\alpha{}}}
  \exp{}\beta{}_{v}\ssum{\alpha{}}H(u,v^{\alpha{}})=  \\
  =\int{}\frac{N\beta{}_{v}dxdy}{4i\pi{}}
  \exp{}
  \biggl\{
  -N
  \bigl[
  \frac{\beta{}_{v}}{2} xy      
  -  \log\left(2\cosh{}[\beta{}_{v}k(g+x/k)]\right){} - \\
  -  k\log{}\left({}2\cosh{}[\beta{}_{v}(h+y)]\right){}
  \bigr]
  \biggr\}.
\end{multline}
The large $\beta{}$ saddle point equations are
\begin{equation}\label{E:j3}
  \begin{aligned}
    \frac{y_{0}}{2} &=
    \tanh{}\left[-\beta{}_{u}\left(g+\frac{x_{0}}{k}\right)\right]{}
    \approx{} \operatorname{sign}{}\left({}-g+\frac{x_{0}}{k}\right){},\\
    \frac{x_{0}}{2k} &=
    \tanh{}\left[\beta{}_{v}\left(h+y_{0}\right)\right]{}
    \approx{} \operatorname{sign}{}\left({}h+y_{0}\right){}.
  \end{aligned}
\end{equation}
Using $\log{}(2\cosh{}\beta{}z) \approx{}\beta{} |z|$ for
$\beta{}\gg{}1$, one can check that 
\begin{equation}\label{E:j4}
  \lim_{\beta{}_{u},\beta{}_{v} \to{}\infty{}} \frac{-1}{N\beta{}_{u}}\log{Z}
  =\min_{u\in{}[-1,1]}
  \left[{}2u\operatorname{sign}{}(2u+h)+gu+h\operatorname{sign}{}(2u+h)\right],
\end{equation}
where the last expression is the outcome of the game if both players
play optimally, since at optimal play $v=\operatorname{sign}{}(2u+h)$.

\vspace{18pt}
\includegraphics{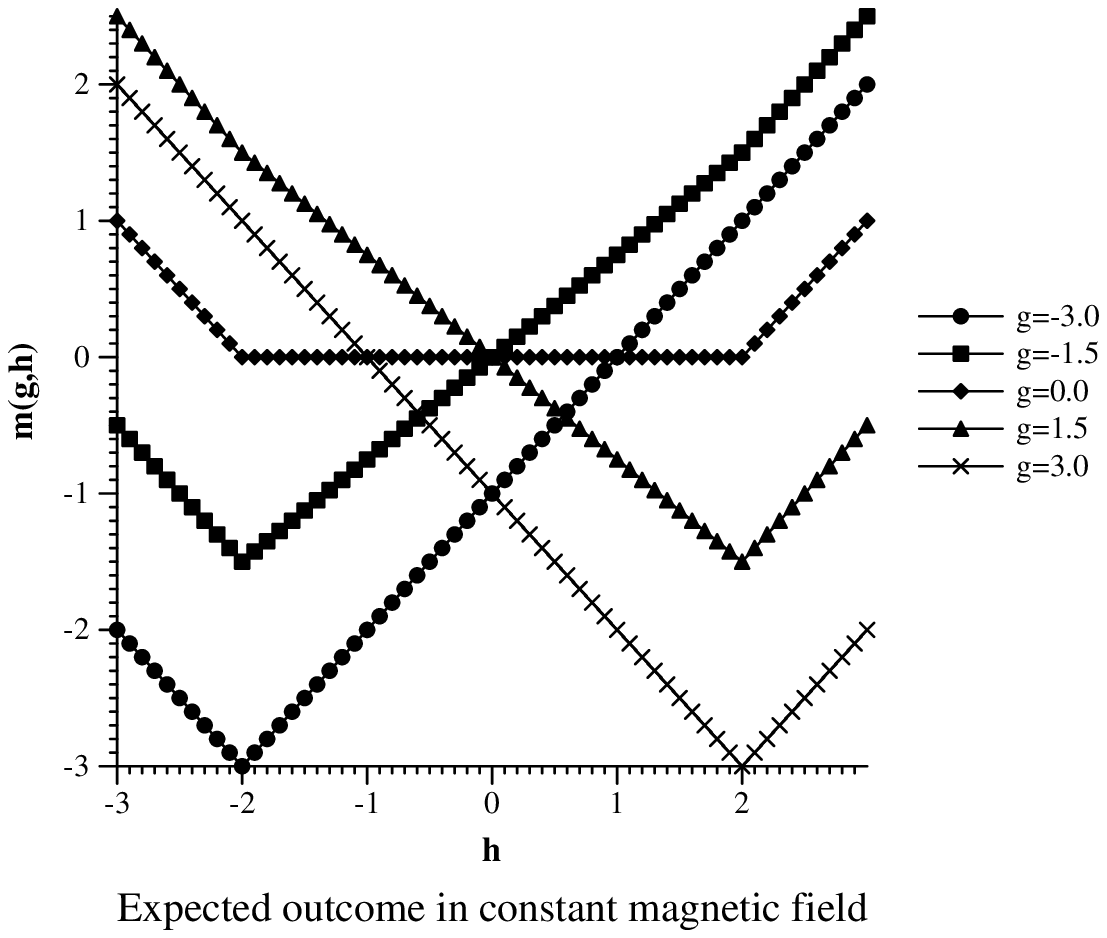}
\vspace{18pt}

In the next example  random magnetic field has been added to the model:
\begin{equation}\label{E:j6}
  H(u,v)=\frac{2}{N}
  \left(
    \ssum{i}u_{i}
  \right)
  \left(
    \ssum{i}v_{i}
  \right)
  +\ssum{i}(g+g_{i})u_{i} + \ssum{i}(h+h_{i})v_{i}, 
\end{equation}
where $\overline{h_{i}}=\overline{g_{i}}=0$ and 
$\overline{h_{i}^{2}}=\overline{g_{i}^{2}}=1$.
After some tedious but standard calculations, we obtain that 
$M(\beta_{u}{},\beta{}_{v})$ is equal to the saddle-point value of
\begin{align}\label{E:j7}
  -\frac{1}{\beta_{u}{}}
  \bigg\{{} 
  \frac{\beta_{u}{}}{2}pq\, +
  &\int{}\frac{dz}{\sqrt{2\pi{}}}e^{-\frac{1}{2}z^{2}}
  \log{}\left({}
    2\cosh{}\left[{}-\beta_{u}{}(\sqrt{J}p+g+z)\right]\right) \\
  -\frac{\beta_{u}{}}{\beta_{v}{}}
  &\int{}\frac{dw}{\sqrt{2\pi{}}}e^{-\frac{1}{2}w^{2}}
  \log{}\left({}
    2\cosh{}\left[{}\beta_{v}{}(\sqrt{J}q+h+w)\right]\right)
\end{align}
with respect to $p$ and $q$.
The saddle point equations are
\begin{align}\label{E:j89}
  p_{0}&=
  2\sqrt{J} \int{}\frac{dw}{\sqrt{2\pi{}}}e^{-\frac{1}{2}w^{2}}
  \tanh{}\left[{}\beta{}_{v}
    (\sqrt{J}q_{0}+h+w)\right]{}  \\
  q_{0}&=
  2\sqrt{J} \int{}\frac{dz}{\sqrt{2\pi{}}}e^{-\frac{1}{2}z^{2}}
  \tanh{}\left[{}-\beta{}_{u}
    (\sqrt{J}p_{0}+g+z)\right]{}.
\end{align}
The numerical solution of these equations is presented on the
following graph:

\vspace{18pt}
\includegraphics{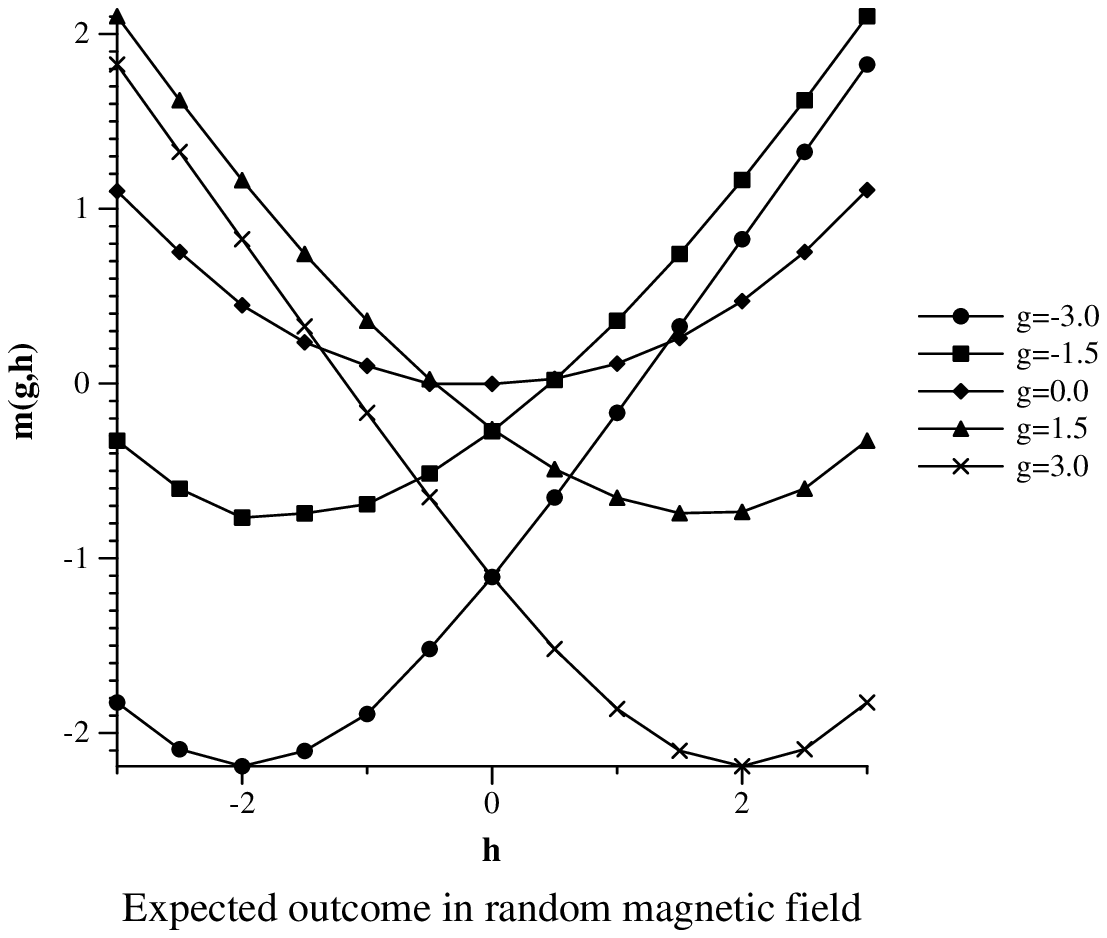}
\vspace{18pt}

Since the Hamiltonian \eqref{E:j6} is fairly simple, the expression  
\eqref{E:j7} can be derived without the use of replicas, too. For that
purpose, we assume that $M(\beta{}_{u},\beta{}_{v})$ receives its 
dominant contribution from spin configurations where the $u$ spins' 
average magnetisation is $\overline{u}$.
Then
\begin{multline}\label{E:j10}
  M(\beta_{u}{},\beta_{v}{})= \\
  =\frac{-1}{\beta_{u}{}N}
  \int{}\prod_{i}\frac{dg_{i}}{\sqrt{2\pi{}}}
  e^{-\left({}\frac{1}{2}\ssum{i}g_{i}^{2}\right){}}
  \log{}\Biggl\{{}\int{}d\lambda{}
  \sum_{\{u_{i}\}}e^{-i\lambda{}\left(\ssum{i}u_{i}-N\bar{u}\right)}{}
  \notag{}          \\
  \exp{}\left(
    -\beta{}_{u}\left[{}
      \ssum{i}
      (g+g_{i})u_{i}+Nf_{v}(\bar{u})\right]{}\right)\Biggr\} \notag{},
\end{multline}
where $f_{v}(\bar{u})$ is the free-energy of the $v_{i}$ spins in the  
external field of the $u_{i}$ spin variables:
\begin{equation}\label{E:j11}
  f_{v}(\bar{u})=
  \frac{1}{\beta_{v}}\int{}\frac{dw}{\sqrt{2\pi{}}}
  e^{-\frac{1}{2}w^{2}}
  \log{}\left({}2\cosh{}
    \left[
      \beta_{v}{}(2J\bar{u}+h+w)
    \right]
  \right){}.
\end{equation}
$M(\beta_{u}{},\beta_{v}{})$ evaluates to 
\begin{equation}\label{E:j12}
  \begin{aligned}
    \frac{-1}{\beta_{u}{}}
    \biggl[
    i\lambda{}\bar{u}+
    \int{}\frac{dz}{\sqrt{2\pi{}}}e^{-\frac{1}{2}z^{2}}
    \log{}\left({}2\cosh{}\left[{}-\beta{}_{u}(g+z)+
        i\lambda{}\right]{}\right)\\
    -\frac{\beta_{u}{}}{\beta_{v}{}}
    \int{}\frac{dw}{\sqrt{2\pi{}}}e^{-\frac{1}{2}w^{2}}
    \log{}\left({}2\cosh{}\left[{}\beta{}_{v}(2J\bar{u}+h+w)\right]{}\right)
    \biggr].
  \end{aligned}
\end{equation}
This formula should be computed at its saddle-point value with respect to 
$\lambda$ and its minimum with respect to $\bar{u}\in{}[-1,1]$. After the
change of variables $\bar{u}=q/(2\sqrt{J})$ and 
$i\lambda{}=\beta_{u}{}\sqrt{J}p$, the expressions \eqref{E:j7} and 
\eqref{E:j12} coincide. Since we managed to evaluate 
$M(\beta_{u}{},\beta_{v}{})$ without the use of replicas, too,
this example is certainly not the most impressive application of
the negative replica method. Nevertheless, this model provides an 
example where one can  analytically prove that the replica method  works.

Finally, we attempt to treat the case of a spin-glass type objective
function  
\begin{equation}\label{E:s1}
  H_{J}(s_{i},h_{i})=\sum_{1\leq{}i\le{}j\leq{}N}J_{ij}s_{i}s_{j}+
  g\sum_{1\leq{}i\leq{}N}h_{i}s_{i},
  \qquad{}h_{i}=\pm{}1,\quad{}s_{i}=\pm{}1,
\end{equation}
where $J_{ij}$ is  random variable with  Gaussian distribution
\begin{equation}
  d\mu{}(J_{ij})=\sqrt{N/2J}\exp{(-J^{2}/2N)}dJ_{ij}.
\end{equation}
The minimizer makes
the first move and controls the $h_{i}$ variables, while the
maximizer makes the second move and controls the $s_{i}$ spins.
The partition function of this system is
\begin{equation}\label{E:s2}\notag{}
  \begin{aligned}
    Z_{n,k}=&\int{}\prod_{i<k}d\mu{}(J_{ik})
    \sum_{\{h_{i}^{a},s_{i}^{a\alpha}\}}
    \exp{}\beta{}_{v}\ssum{a\alpha{}}H_{J}(h_{i}^{a},s_{i}^{a\alpha{}})\\
    =&\int{}\prod_{a\alpha{}<{}b\beta{}}
    \bigl({}
    \sqrt{\frac{N\beta{}}{2\pi{}}}
    dQ_{a\alpha{}b\beta{}}\bigr){}
    \exp -N
    \Biggl\{{}
    -nk\frac{\beta{}_{v}^{2}}{4}
    +\frac{\beta_{v}^{2}{}}{2}\sum_{a\alpha{}<{}b\beta{}}
    Q_{a\alpha{}b\beta{}}^{2}-\\  
    &\phantom{==}-\log{}\sum_{\{S^{a\alpha{}},H^{a}\}}
    \exp{}\beta{}_{v}
    \biggl[{}
    \beta_{v}{}\sum_{a\alpha{}<{}b\beta{}}
    Q_{a\alpha{}b\beta{}}S^{a\alpha}S^{b\beta}{}-
    g\sum_{a}H^{a}\sum_{\alpha}S^{a\alpha}
    \biggr]
    \Biggr\},
  \end{aligned}
\end{equation}
where $k=-\beta_{u}{}/\beta_{v}{}$.
In the replica symmetric approximation
\begin{equation}\notag{}
  Q_{a\alpha{}a\beta{}}=p,\,Q_{a\alpha{}b\beta{}}=q\quad{}
  \text{for}\quad{} a\neq{}b,
\end{equation}
$Z_{n,k}$ equals to
\begin{equation}\label{E:s3}\notag{}
  \begin{aligned}
    Z&_{n,k}=\\
    \int{}&\prod_{a\alpha{}<{}b\beta{}}
    \bigl({}\sqrt{\frac{N\beta{}}{2\pi{}{}}}dQ_{a\alpha{}b\beta{}}\bigr){}
    \exp{}-N\Biggl\{{}
    \beta{}_{v}^{2}
    \biggl({}
    -\frac{nk}{4}+\frac{n(n-1)k^{2}}{4}q^{2}+\frac{nk(k-1)}{4}p^{2}+
    \frac{nk}{2}p
    \biggr){}-\\
    &-\log{}
    \int\frac{dx}{\sqrt{2\pi{}}}e^{-\frac{1}{2}x^{2}}
    \biggl(
    \int{}\frac{dy}{\sqrt{2\pi{}}}e^{-\frac{1}{2}y^{2}}\times{}\\
    &\times{}\biggl\{
    \bigl[{}2\cosh{}(\beta_{v}{}(\sqrt{q}x+\sqrt{p-q}y+g))\bigr]{}^{k}+
    \bigl[{}2\cosh{}(\beta_{v}{}(\sqrt{q}x+\sqrt{p-q}y-g))\bigr]{}^{k}
    \biggr\}
    \biggr){}^{n}
    \Biggr\}.
  \end{aligned}
\end{equation}
From this equation one obtains the expected outcome of the game:
\begin{equation}
  \begin{aligned}\notag{}
    m&=\frac{1}{N}\min_{\{h_{i}\}}
    \bigl({}\max_{\{s_{i}\}}H_{J}(s_{i},h_{i})\bigr){}
    =\lim_{n\to{}0,\,\beta_{v}\to{}\infty{}}\frac{1}{k\beta{}_{v}nN}
    (Z_{n,k}-1)=\notag{}\\
    &=\frac{\beta{}_{v}}{4}
    \left({}1+kq^{2}+(1-k)p^{2}-2p\right){} +\\
    &+\frac{1}{k\beta{}_{v}}
    \int{}\frac{dx}{\sqrt{2\pi{}}}
    e^{-\frac{1}{2}x^{2}}\log{}\int{}\frac{dy}{\sqrt{2\pi{}}}
    e^{-\frac{1}{2}y^{2}}
    \biggl\{
    \bigl[{}
    2\cosh{}(\beta{}_{v}(\sqrt{q}x+\sqrt{p-q}y+g))
    \bigr]{}^{k}\\
    &\phantom{==============}+\bigl[{}
    2\cosh{}(\beta{}_{v}(\sqrt{q}x+\sqrt{p-q}y-g))
    \bigr]{}^{k}
    \biggr\}
  \end{aligned}
\end{equation}
where the last expression should be evaluated at its saddle
point. This expression is very similar to the free energy of a spin
glass at the one stage replica symmetry breaking approximation
\cite{parisi2}. Indeed, $Q_{a\alpha{}b\beta}{}$ might be regarded  as
a $nk\times{}nk$ matrix broken into blocks of size
$k\times{}k$. However, here $k$ is a fixed negative
number. 
$m(p,q)$  has a
minimum at $p=q=1$ on the line $p=q$.
In this approximation
\begin{equation}
  m_{minimax}(g)=\int{}\frac{dx}{\sqrt{2\pi{}}}
  e^{-\frac{1}{2}x^{2}}
  \min{(|x+g|,|x-g|)}.
\end{equation}
A better approximation is achived if we search for the saddle point on
the $(p,q)$ plane. Since the first term of $m$ scales as $O(\beta{})$
as $\beta{}\to{}\infty{}$, while the second has finit limes, the
saddle point should be on the curve $1+kq^{2}+(1-k)p^{2}-2p$.
We evaluated numerically $m$ as the function of $g$.
We plot the function $m_{minimax}(g)$ (solid line on the figure).

\vspace{18pt}
\includegraphics{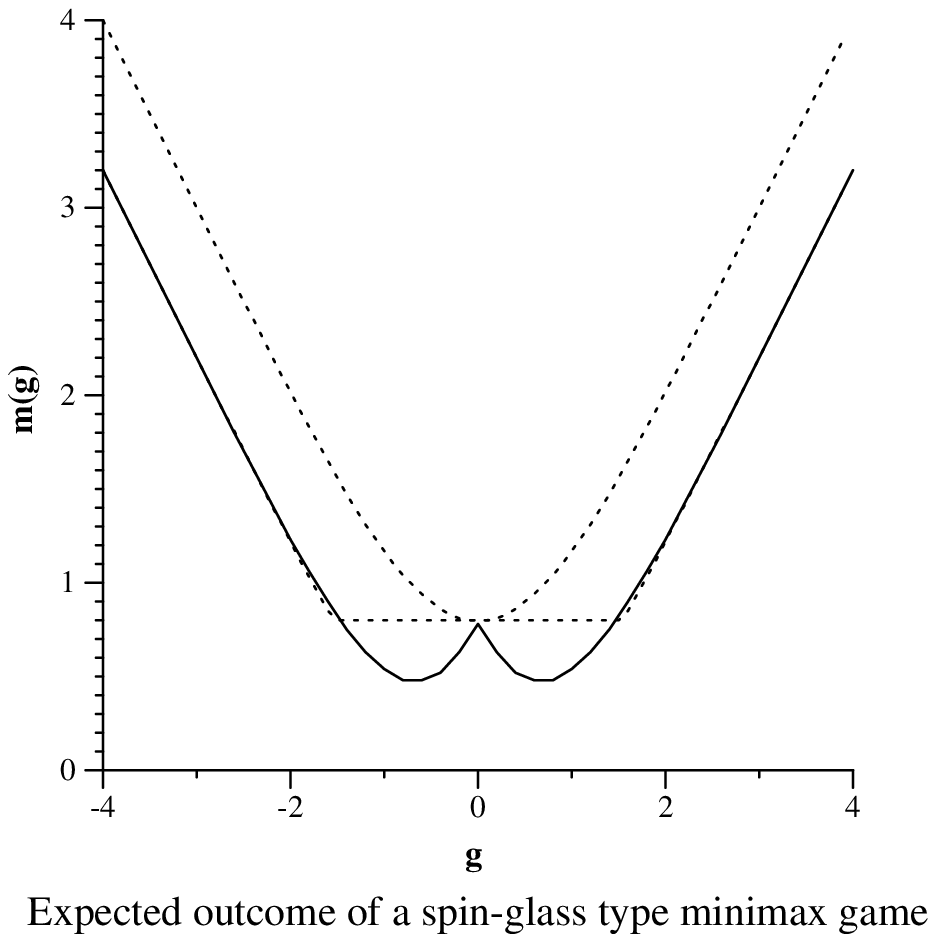}
\vspace{18pt}

The exact value of
$m_{minimax}(g)$ is smaller than $m_{spinglass}(g)$,
(where $m_{spinglass}(g)$ is the maximal value of
 $\sum{}J_{ij}s_{i}s_{j}+g\sum{}s_{i}$),
  since
the minimizer tries to set $h_{i}$ into directions least favorable for
the maximizer, while the constant magnetic field is equivalent to a
randomly choosen $h_{i}$ configuration. However, 
$m_{minimax}(g)\geq{}(m_{spinglass}(0))$, since one of
$\pm{}\sum{}h_{i}s_{i}$ is always nonnegative. 
 $m_{minimax}(g)\geq{}(g-m_{spinglass}(0))$ also holds, since if the
 spins are set to the same direction as $h_{i}$, then the contribution
 of $\sum{}J_{ij}s_{i}s_{j}$ cannot be less than $-m_{spinglass}(0)$
by the symmetry of the couplings $J_{ij}$. We also expect that 
$m_{minimax}(g)$ converges to $g-m_{spinglass}(0)$ as
$g\to{}\infty{}$.
These considerations provide upper and lower bounds for
$m_{minimax}(g)$
(dotted lines on the figure).
Unfortunately, the lower bound is violated 
 for small $g$, while 
its assimptotics is correctly reproduced.
 It would be interesting to know if a
better, replica symmetry breaking solution  would cure this problem.

\section{Multi-move games}

Up to this point only two-moves games were treated. The extension for
multi-move games is straightforward. For example, the outcome of the
four-move game
\[
M=\inf_{u}\left\{\sup_{v}\left[\inf_{w}\left(\sup_{z}
H(u,v,w,z)\right)\right]\right\}
\]
is
\[
\lim_{\beta_{u,v,w,z} \to \infty}
\frac{-1}{\beta_{u}}\lim_{n \to 0}
\left[
\left(
\sum_{\{u^{a},v^{a\alpha},w^{a,\alpha\beta},z^{a\alpha\beta\gamma}\}}
\exp \beta_{z} \sum_{a\alpha\beta\gamma}
H(u^{a},v^{a\alpha},w^{a\alpha\beta},z^{a\alpha\beta\gamma})
\right)-1
\right]
\]
where the ranges of the indices are
$|a|=n,\,|\alpha{}|=-\beta{}_{u}/\beta{}_{v},\,
|\beta{}|=-\beta{}_{v}/\beta{}_{w}\,|\gamma{}|=-\beta{}_{w}/\beta{}_{z}$.

The limit $\beta_{u,v,w,z} \to \infty$ corresponds to the optimal
strategies of the players. Finite $\beta$ simulates non-exact
optimization, i.e. players with bounded computational capabilities. An
interesting case is when one player's temperature is infinite, so the
other ones play against random moves. Such games are called 'games
against Nature' \cite{against}.

\section{Discussion}

In the previous sections we used  the method of negative replicas 
to examine optimization problems arising in minimax games.  Such games provide
examples of very difficult combinatorial problems. In principle
our method is able to estimate the expected outcome of some adversary
games. Unfortunately, due to the complexity of the calculations emerging 
in problems of spin glass type, we manage to treat only fairly simple
 optimization problems. Nevertheless, the method of negative replicas provides
a natural framework to treat game theoretical problems with the
machinery of statistical mechanics.

\end{document}